\documentclass{desyproc}

\IfFileExists{srcltx.sty}{\usepackage[active]{srcltx}}

\usepackage{xspace} %
\usepackage[sort&compress,numbers]{natbib}

\newcommand{\lya}{Ly-$\alpha$\xspace}
\newcommand{\kev}{\:\mathrm{keV}} 
\newcommand{\dm}{\textsc{dm}}%
\newcommand{\dw}{{\textsc{nrp}}} 

\begin{document}
\title{Bounds on Light Dark Matter}

\author{{\slshape %
    Alexey Boyarsky$^{1,2}$, Oleg Ruchayskiy$^3$}\\[1ex]
  $^{1}$ETHZ, Z\"urich, CH-8093,
  Switzerland\\
  $^{2}$Bogolyubov Institute for Theoretical Physics, Kiev 03780, Ukraine\\
  $^{3}$\'EPFL, FSB/ITP/LPPC, BSP 726, CH-1015, Lausanne, Switzerland }

\contribID{lindner\_axel}

\desyproc{DESY-PROC-2008-02}
\acronym{Patras 2008} 
\doi  

\maketitle

\begin{abstract}
  In this talk we review existing cosmological and astrophysical bounds on
  light (with the mass in keV -- MeV range) and super-weakly interacting
  dark matter candidates. A particular attention is paid to the sterile
  neutrino DM candidate.
\end{abstract}

The nature of Dark Matter (DM) is one of the most intriguing questions of
modern physics. Its resolution would have a profound impact on the development
of particle physics beyond its Standard Model (SM).  Although the possibility
of having massive compact halo objects (MACHOs) as a dominant form of DM is
still under debates (see recent discussion in~\cite{Calchi:07} and references
therein), it is widely believed that DM is made of non-baryonic particles.
Yet the SM of elementary particles does not contain a viable DM particle
candidate -- massive, neutral and long-lived particle.  Active neutrinos,
which are both neutral and stable, form structures in a top-down
fashion~\cite{Zeldovich:70}, and thus cannot produce observed large scale
structure.  Therefore, the DM particle hypothesis implies the extension of the
SM. Thus, constraining properties of the DM, helps to distinguish between
various DM candidates and may help to differentiate among different beyond the
SM models (BSM).  \emph{What is known about the properties of DM particles?}

\paragraph{A lower bound on the mass of DM particle.}
\label{sec:lower-bound-mass}

The DM particle candidates have very different masses (for reviews see
e.g.~\cite{Bergstrom:00,Bertone:04}). Quite a robust and model-independent
\emph{lower bound} on the mass of DM particles was suggested
in~\cite{Tremaine:79}. The idea was based on the fact that for any fermionic
DM the average phase-space density (in a given DM-dominated, gravitationally
bound object) cannot exceed the phase-space density of the degenerate Fermi
gas.  This argument, applied to the most DM-dominated dwarf spheroidal
satellites (dSph's) of the Milky Ways leads to the bound $m_\dm >
0.41\kev$~\cite{Boyarsky:08a,Gorbunov:08b}.

For particular DM models (with the known primordial velocity dispersion) and
under certain assumptions about the evolution of the system which led to the
observed final state, this limit can be strengthened.  This idea was developed
in a number of works (see e.g. refs. in~\cite{Boyarsky:08a}).

\paragraph{Decaying DM.}
\label{sec:decaying-dm}

For any DM candidate there should exist a mechanism of its production in the
early Universe. Although it is possible that the DM is produced through
interactions with non-SM particles only (e.g. from inflaton decay) and is
inert with respect to all SM interactions, many viable DM candidates are
produced via interaction with the SM sector. According to this interaction the
DM candidates can be subdivided into \emph{annihilating} and \emph{decaying}
ones. The annihilating DM candidates -- WIMPs~\cite{Lee:77} -- are well
studied. A decaying DM candidate should be \emph{superweakly} interacting
(i.e. weaker than electroweak), otherwise it cannot have a cosmologically long
lifetime.  There are many examples of \emph{super-WIMP} DM models: sterile
neutrinos~\cite{Dodelson:93}, gravitino in theories with broken
R-parity~\cite{Takayama:00,Buchmuller:07}, light volume
modulus~\cite{Conlon:07}, Majoron~\cite{Lattanzi:07}. All these candidate
posses a 2-body decay channel: $\dm \to \gamma + \nu,\gamma+\gamma$.
Therefore, searching for a monochromatic decay line in the spectra of
DM-dominated objects provides a way of \emph{indirect detection} of the DM or
helps to \emph{constrain} its interaction strength with the SM particles.

The astrophysical search for \emph{decaying} DM is promising and a positive
result would be much more conclusive, than in the case of annihilating DM.
Indeed, the decay signal is proportional to the \emph{column density}:
$\int\rho_\dm(r)dr$ along the line of sight and not to the
$\int\rho^2_\dm(r)dr$ (as it is the case of the annihilating DM). As a result
\emph{(i)} a vast variety of astrophysical objects of different nature would
produce roughly the same decay signal~\cite{Boyarsky:06c,Bertone:07};
\emph{(ii)} this gives a freedom of choosing the observational targets,
allowing to avoid the complicated astrophysical backgrounds (e.g. one does not
need to look at the Galactic center, expecting a comparable signal from dark
outskirts of galaxies, clusters and dark dSphs); \emph{(iii)} if a candidate
line is found, its surface brightness profile may be measured (as it does not
decay quickly away from the centers of the objects), distinguished from
astrophysical lines (which usually decay in outskirts) and compared among
several objects with the same expected signal.  This makes astrophysical
search for decaying DM \emph{another type of a direct detection experiment}.

\begin{figure}[h]
  \centerline{\includegraphics[width=\textwidth]{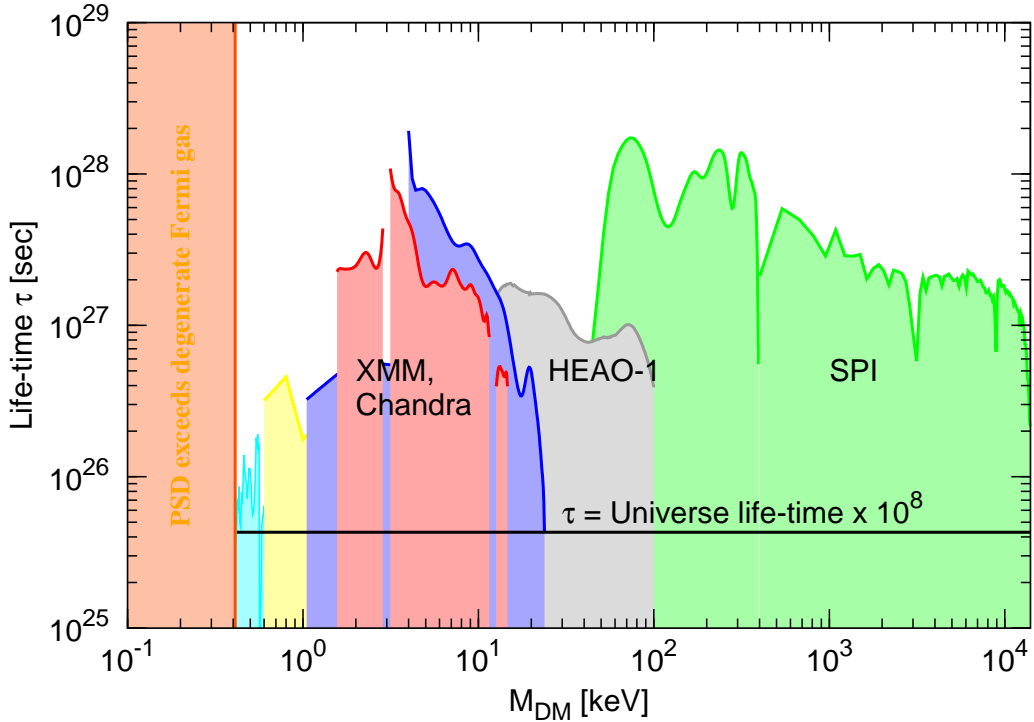}}
  \caption{Restrictions on the lifetime of the radiatively decaying DM (based
    on~\cite{Boyarsky:05,Boyarsky:06c}).  The lifetime exceeds the age of the
    Universe by at least $10^8$.}\label{Fig:decay}
\end{figure}

A search of the DM decay signal was conducted both in the keV -- MeV
range~\cite{Boyarsky:05,Boyarsky:06b,Boyarsky:06c,Riemer:06,Watson:06,Boyarsky:06e,Abazajian:06b,Boyarsky:06d,Boyarsky:06f,Boyarsky:07a,Boyarsky:07b}
and in GeV range~\cite{Bertone:07}. The aggregate constraints on the decaying
DM lifetime (towards the radiative decay) are shown on Fig.~\ref{Fig:decay}

\paragraph{\lya constraints.}
\label{sec:cosm-constr}

The fable strength of interaction of light super-WIMP particles often means
that they were produced in the early Universe in a non-thermal way and
decoupled deep into the radiation dominated (RD) epoch, while still being
relativistic. This makes these particles \emph{warm DM} candidates (WDM) (see
e.g.~\cite{Bode:00}).

An important way to distinguish between WDM and CDM models is the analysis of
the Lyman-$\alpha$ (Ly-$\alpha$) forest data~(for an introduction see
e.g.~\cite{Hui:97,Gnedin:01,Weinberg:03}).  Although very promising, the
Ly-$\alpha$ method is very complicated and indirect. As at redshifts, probed
by Ly-$\alpha$, the evolution of structure already enters a non-linear stage,
to relate measured power spectrum with the parameters of each cosmological
model, one would have to perform prohibitively large number of numerical
simulations.  Therefore, various simplifying approximations have to be
realized (see e.g.~\cite{McDonald:05}).  Apart from computational
difficulties, the physics, entering the \lya analysis is not fully understood,
as it is complicated and can be significantly influenced by DM particles~(see
e.g.~\cite{Viel:2003fx,Kim:07a,Bolton:07a,Gao:07,Stasielak:06,Faucher:07}).
Bayesian approach, used to fit the cosmological data, should also be applied
with caution to put bounds on the particle physics
parameters~\cite{Boyarsky:08b}.

In many super-weakly interacting DM models, due to the non-thermal primordial
velocity distribution, the linear powerspectrum (PS) (used as initial
conditions in \lya analysis) has complicated non-universal form. The analysis
of~\cite{Viel:06,Seljak:06,Viel:07} assumed PS with a cut-off at small scales,
defined by the particle's velocities. These results are not applicable for
many models of decaying DM. For example, in a number of models (sterile
neutrinos, gravitino) the primordial velocity distribution is a mixture of
colder and warmer components and the PS develops a plateau at small scales.
This makes much smaller masses compatible with \lya
bounds. 
For these smaller masses it is important to take into account explicitly the
primordial velocities of the particles (and not only their effect on the PS).
See detailed analysis~\cite{Boyarsky:08b}.


\paragraph{Sterile neutrino DM.}
\label{sec:sterile-neutrino-dm}

Although known as a DM candidate for some 15 years~\cite{Dodelson:93}, the
sterile neutrino DM recently attracted a lot of attention.  It was
shown~\cite{Asaka:05a,Asaka:05b} that if one adds three right-handed (sterile)
neutrinos to the SM, it is possible to explain simultaneously the data on
neutrino oscillations, the DM in the Universe and generate the correct baryon
asymmetry of the Universe without introducing any new physics \emph{above
  electro-weak scale}. The lightest (DM) sterile neutrino can have mass in
keV-MeV range and be coupled to the rest of the matter weakly enough to
provide a viable (\emph{cold} or \emph{warm)} DM candidate.  This model,
explaining the three observed BSM phenomena within one consistent framework,
is called \emph{the $\nu$MSM}~\cite{Asaka:05a,Shaposhnikov:07b}.

There are several
mechanisms of production of DM sterile neutrino in the early Universe: non-resonant active-sterile neutrino oscillations
(\textbf{NRP})
~\cite{Dodelson:93,Asaka:06c},
resonant oscillations in the presence of lepton asymmetry
(\textbf{RP})~\cite{Shi:98,Shaposhnikov:08a}, decay of the gauge-singlet
scalar field~\cite{Shaposhnikov:06} (see also \cite{Kusenko:06a,Petraki:08}).
The \lya analysis was performed so far only for NRP scenario, and the results
were claimed to be in the range $5-15\kev$ (see also~\cite{Boyarsky:08b}).
Phase-space density bounds, applied to the NRP scenario lead to the $m_\dw >
1.7 - 4\kev$.

\begin{figure}
  \begin{tabular}[c]{cc}
    \includegraphics[width=.5\linewidth]{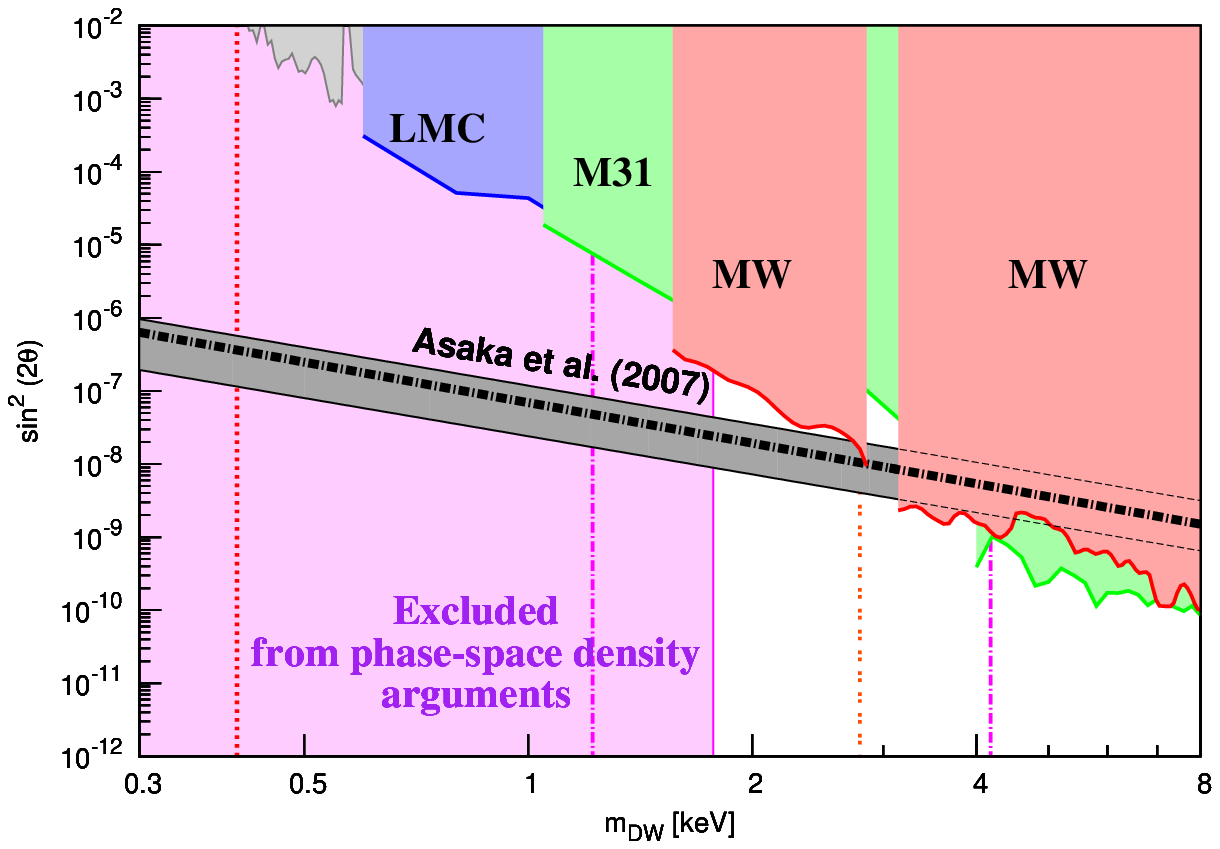} &
    \includegraphics[width=.5\linewidth]{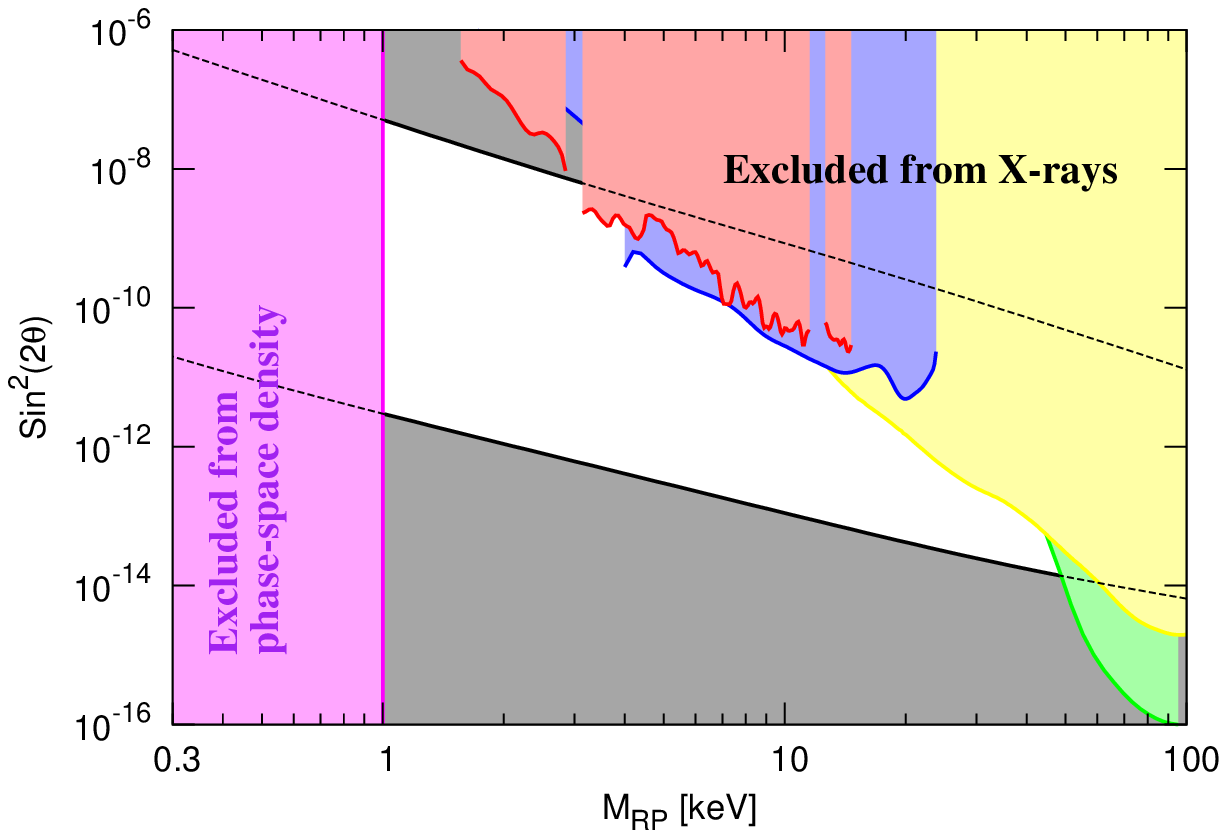} \\
    \textbf{(a)} & \textbf{(b)}
  \end{tabular}
  \caption{Restrictions on sterile neutrino DM in NRP (left) and RP (right)
    scenarios.}
  \label{fig:sterile}
\end{figure}

Combining various constraints we see that there is a tension between the NRP
scenario and the data (X-ray bounds and phase-space density arguments). For
the RP mechanism a large window of allowed parameters remain open. These
results are summarized on Fig.~\ref{fig:sterile}

\begin{footnotesize}
  \setlength{\bibsep}{0.0pt}

\let\jnlstyle=\rm\def\jref#1{{\jnlstyle#1}}\def\aj{\jref{AJ}}
  \def\araa{\jref{ARA\&A}} \def\apj{\jref{ApJ}} \def\apjl{\jref{ApJ}}
  \def\apjs{\jref{ApJS}} \def\ao{\jref{Appl.~Opt.}} \def\apss{\jref{Ap\&SS}}
  \def\aap{\jref{A\&A}} \def\aapr{\jref{A\&A~Rev.}} \def\aaps{\jref{A\&AS}}
  \def\azh{\jref{AZh}} \def\baas{\jref{BAAS}} \def\jrasc{\jref{JRASC}}
  \def\memras{\jref{MmRAS}} \def\mnras{\jref{MNRAS}}
  \def\pra{\jref{Phys.~Rev.~A}} \def\prb{\jref{Phys.~Rev.~B}}
  \def\prc{\jref{Phys.~Rev.~C}} \def\prd{\jref{Phys.~Rev.~D}}
  \def\pre{\jref{Phys.~Rev.~E}} \def\prl{\jref{Phys.~Rev.~Lett.}}
  \def\pasp{\jref{PASP}} \def\pasj{\jref{PASJ}} \def\qjras{\jref{QJRAS}}
  \def\skytel{\jref{S\&T}} \def\solphys{\jref{Sol.~Phys.}}
  \def\sovast{\jref{Soviet~Ast.}} \def\ssr{\jref{Space~Sci.~Rev.}}
  \def\zap{\jref{ZAp}} \def\nat{\jref{Nature}} \def\iaucirc{\jref{IAU~Circ.}}
  \def\aplett{\jref{Astrophys.~Lett.}}
  \def\apspr{\jref{Astrophys.~Space~Phys.~Res.}}
  \def\bain{\jref{Bull.~Astron.~Inst.~Netherlands}}
  \def\fcp{\jref{Fund.~Cosmic~Phys.}} \def\gca{\jref{Geochim.~Cosmochim.~Acta}}
  \def\grl{\jref{Geophys.~Res.~Lett.}} \def\jcp{\jref{J.~Chem.~Phys.}}
  \def\jgr{\jref{J.~Geophys.~Res.}}
  \def\jqsrt{\jref{J.~Quant.~Spec.~Radiat.~Transf.}}
  \def\memsai{\jref{Mem.~Soc.~Astron.~Italiana}}
  \def\nphysa{\jref{Nucl.~Phys.~A}} \def\physrep{\jref{Phys.~Rep.}}
  \def\physscr{\jref{Phys.~Scr}} \def\planss{\jref{Planet.~Space~Sci.}}
  \def\procspie{\jref{Proc.~SPIE}} \let\astap=\aap \let\apjlett=\apjl
  \let\apjsupp=\apjs \let\applopt=\ao

\end{footnotesize}

\end{document}